\documentclass[a4paper]{jpconf}

\usepackage{graphicx}

\begin{document}

\begin{flushright}
{\large \tt TTK-12-37}
\end{flushright}

\title{Triplet seesaw model: from inflation to asymmetric dark matter and leptogenesis}

\author{Chiara Arina}

\address{Institut f\"ur Theoretische Teilchenphysik und Kosmologie, RWTH Aachen, 52056 Aachen, Germany}

\ead{chiara.arina@physik.rwth-aachen.de}

\begin{abstract}
The nature of dark matter (DM) particles and the mechanism that provides their measured relic abundance are currently unknown. Likewise, the nature of the inflaton is unknown as well. We investigate the triplet seesaw model in an unified picture. At high energy scale, we consider Higgs inflation driven by an admixture of standard model and triplet Higgs fields, both coupled non-minimally to gravity. At intermediate and low energies we investigate vector like fermion doublet DM candidates with a charge asymmetry in the dark sector, which is generated by the same mechanism that provides the baryon asymmetry, namely baryogenesis-via-leptogenesis induced by the decay of scalar triplets. At the same time the model gives rise to neutrino masses in the ballpark of oscillation experiments via type-II seesaw. We then apply Bayesian statistics to infer the model parameters giving rise to the observed baryon asymmetry and DM density, compatibly with inflationary and DM direct detection constraints, updated with the CRESST-II excess, the new XENON100 data release and KIMS exclusion limit.
\end{abstract}

\section{Introduction}

A widely accepted theory describing the early universe is inflation and many realization of it have been modeled (see {\it e.g.}~\cite{Lyth:1998xn,Martin:2006rs} for a review). Among them, the Higgs boson has been suggested as inflaton~\cite{Bezrukov:2007ep} by means of a non-minimal coupling to gravity $\xi_H$: inflation will occur at the unitarity scale $M_{\rm pl}/\xi_H \simeq 10^{14}$ GeV. The observation of a new boson at LHC by CMS~\cite{:2012gu} and ATLAS~\cite{:2012gk} points towards a Higgs-like particle with a mass of 125 GeV. At this mass value the Higgs potential is likely in a metastable vacuum~\cite{Degrassi:2012ry} around $10^9$ GeV. A viable way of curing the potential instability  and restoring perturbative unitarity is to introduce new physics at that scale, {\it i.e.} a triplet scalar with a mass $M_\Delta = 10^8$ GeV, as it has been discussed in~\cite{Arina:2012fb}. The triplet mixes with the Higgs and has a non-minimal coupling to the Ricci scalar $R$, $\xi_\Delta$, such that both scalars behave as inflaton: the potential will be flat enough to produce 60 $e$-folds from slow-roll needed to solve the horizon problem. At the end of inflation the early universe enters the radiation era: within the thermal bath the visible and dark matter are created in the measured amount~\cite{Komatsu:2010fb}: $\Omega_{\rm B} \simeq 0.04$ and $\Omega_{\rm DM} \simeq 0.23$. We suppose that the DM is asymmetric as the baryonic matter, instead its relic abundance being due to the usual freeze-out mechanism. Both asymmetries are linked together and induced by triplet decay via leptogenesis, as detailed in~\cite{Arina:2011cu,Arina:2012fb}. The DM candidate is part of a doublet under $SU(2)_L$, either scalar or fermion, and leads to distinct signatures in DM direct detection experiments: its scattering is inelastic~\cite{TuckerSmith:2001hy} and accounts for the DAMA/LIBRA signal~\cite{Bernabei:2010mq}. We investigate if it may account for the CRESST-II excess~\cite{Angloher:2011uu} as well, while being compatible with the most recent upper bound by XENON100~\cite{:2012nq} and by KIMS~\cite{Kim:2012rz}.

\section{The model and the DM candidate}\label{sec:model}

The standard model (SM) is extended by introducing a scalar triplet $\Delta$  with charges $(3,2)$ under the gauge group $SU(2)_L \times U(1)_Y$. The scalar potential involving the Higgs field as well is:
\begin{eqnarray}\label{eq:ScalarPotential}
V_J(\Delta, H)  & = &  M_\Delta^2 \Delta^\dagger \Delta + \frac{\lambda_\Delta}{2} (\Delta^\dagger \Delta)^2 -  M_H^2 H^\dagger H + \frac{\lambda_H}{2} (H^\dagger H)^2\nonumber\\ & + &  \lambda_{\Delta H} H^{\dagger} H \Delta^\dagger \Delta + \frac{1}{\sqrt{2}} \left[ \mu_H \Delta^\dagger H H + {\rm h.c.} \right]  \,.
\end{eqnarray}
The bilinear couplings of the triplet with the leptons are:
\begin{equation}\label{eq:Lag-CP}
 -\mathcal{L}  \supset  +  \frac{1}{\sqrt{2}} \left[ f_H \Delta^\dagger H H + f_L \Delta L L + {\rm h.c.} \right]\,,
\end{equation}
with the redefinition $f_H=\mu_H/M_\Delta$. These terms give rise to neutrino masses via type-II seesaw~\cite{Ma:2006km}, $M_{\nu} \propto f_H f_L (-v^2)/M_\Delta$, where $v$ is the Higgs vev (vacuum expectation value). In order to have Majorana masses in the range indicated by neutrino oscillations, $f_L \simeq 10^{-9}$, because $M_\Delta$ is at the scale of $10^8$ GeV, slightly below the metastability one.

We extend the Lagrangian with the DM candidate given by the neutral component of $\chi \equiv
(\chi^+ \chi^0)^T$ for scalars and of vector-like doublet $\psi \equiv (\psi_{\rm DM}, \psi_-)$ for fermions:
\begin{eqnarray}
- \mathcal{L}_{\rm DM} \supset  
\left\{
\begin{array}{ll}
M_\chi^2 \chi^\dagger \chi + \lambda_\chi (\chi^\dagger\chi)^2 + \left[ \mu_\chi \Delta^\dagger \chi \chi + {\rm h.c.}\right] + & {\rm scalar \, DM} \\
+ \lambda_3 |H|^2 |\chi|^2 + \lambda_4 |H^\dagger \chi|^2
+ \frac{\lambda_5}{2} \left[ (H^\dagger \chi)^2 + {\rm h.c.} \right]\,, & \\
\overline{\psi} i\gamma^\mu \mathcal{D}_\mu \psi + M_D \overline{\psi} \psi  +  \frac{1}{\sqrt{2}} \left[ f_\psi \Delta \psi \psi + {\rm h.c.} \right]\,. & {\rm fermionic \, DM}
\end{array}
\right.
\label{eq:Lag-DM}
\end{eqnarray}
The doublets  are odd under a $Z_2$ symmetry making the DM particle stable, besides being inert, since there are no coupling with the SM fermions. In the case of scalar DM, the $Z_2$ symmetry is elevated to global $U(1)$ Peccei-Quinn symmetry for $\lambda_5 \to 0$, that is null DM number violating coupling. For $\lambda_5\simeq 10^{-7}$ a small mass splitting, $\Delta m \equiv \delta$, is generated between the odd and even neutral states, the lightest one being the DM particle, making the interaction with $Z$ boson off-diagonal. As for the fermionic DM candidate, in the same way a Majorana mass is generated for neutrino masses, the interaction with the triplet scalar generates a Majorana mass for $\psi_{\rm DM}$, $m = f_H f_\psi \frac{-v^2}{M_\Delta}$, via the low scale dimension five operator ${\cal O}_5=\psi\psi H H$, leading to an off diagonal coupling between $\psi_{\rm DM}$ and the $Z$ boson as well. 

\begin{figure}[t]
\begin{minipage}[t]{0.5\textwidth}
\centering
\includegraphics[width=0.97\columnwidth,trim=70mm 3mm 65mm 5mm, clip]{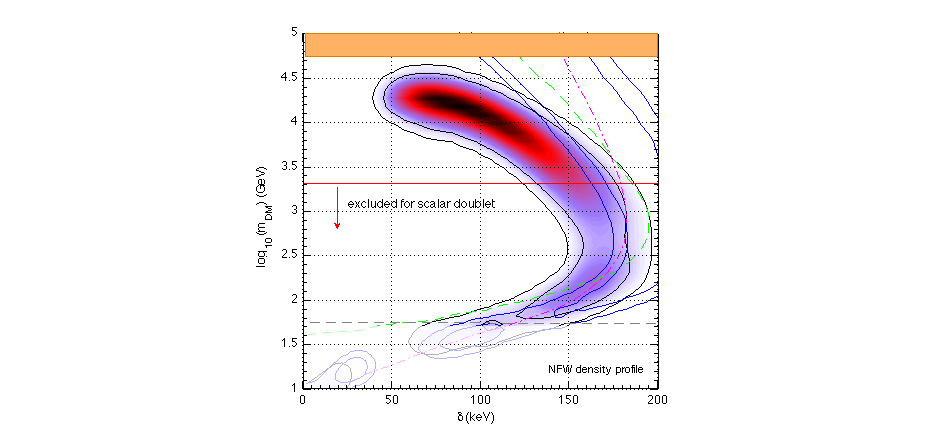}
\caption{2D credible regions in the plane $\{\delta-m_{\rm DM}\}$: individual experimental bounds and regions assuming the NFW DM density profile and marginalizing over the astrophysical uncertainties, combined in a single plot. For DAMA (shaded) and CRESST-II (solid blue)  we show the 90\% and 99\% contours. The $90_S\%$  contours are given by respectively the pink dot-dash curve for XENON100 ($\Delta \chi^2 = 3.1$) and the dashed green line for KIMS ($\Delta \chi^2=4.6$). The region below $m_{\rm DM} =$ 2 TeV is excluded by $\chi_0-\bar{\chi}_0$ oscillation (horizontal red solid line). The light gray line is the $Z$ invisible decay width bound. The orange/dark gray region is excluded by unitarity bound. \label{fig1}}
\end{minipage}
\hspace*{0.2cm}
\begin{minipage}[t]{0.5\textwidth}
\centering
\includegraphics[width=1.09\columnwidth,trim=40mm 92mm 30mm 90mm, clip]{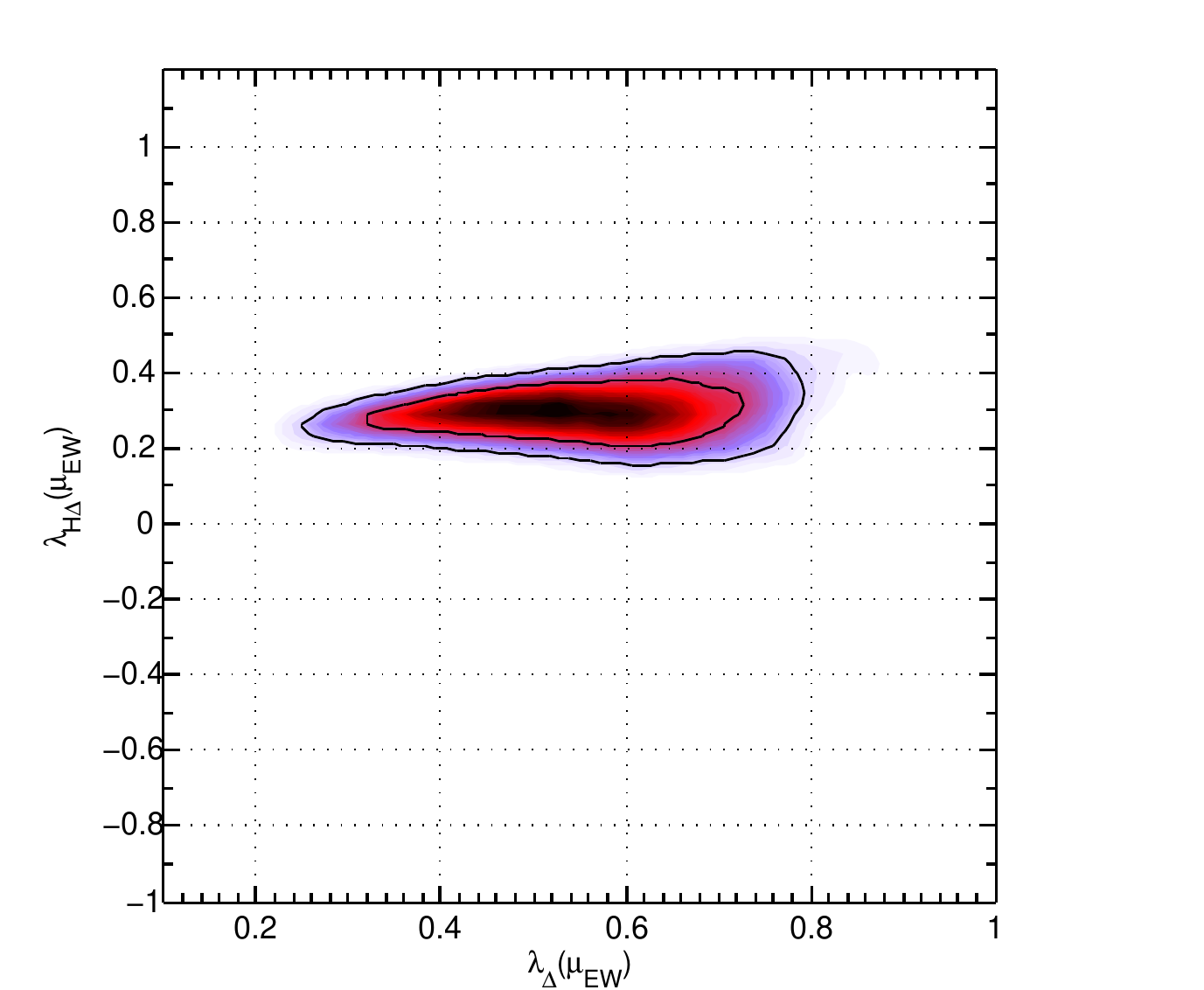}
\caption{2D marginal posterior in the $\{\lambda_{\Delta},\lambda_{H\Delta}\}$-plane for mixed inflation at the EW renormalization scale $\mu_{\rm EW}$. The black solid lines enclose the 68\% and 95\% credible regions. All of other parameters entering in the renormalization group equations, specifically $\xi_H$, $\xi_\Delta$, $\mu_H/M_\Delta$, $f_H$ and $f_\psi$, have been marginalized over. The triplet mass is fixed at $M_\Delta = 10^8$ GeV, while the Higgs mass is set at $M_H=125$ GeV. Predictions for the pure Higgs and pure triplet scenarios are very similar. See for details~\cite{Arina:2012fb} as well as for the Markov Chain Monte Carlo (MCMC) setup.\label{fig2}}
\end{minipage}
\end{figure}

In case of a scalar DM $\chi_0$ any asymmetry in the DM sector gets washed out below electroweak (EW) phase transition by fast oscillations between $\chi_0$ and its complex conjugate field $\overline{\chi_0}$. This sets a limit of the mass scale of $\chi_0$ to be $M_{\chi_0} \equiv m_{\rm DM} \geq 2$ TeV, so that the DM freezes out before oscillations begin to occur.

The DM scattering off nuclei is mediated by the Z boson, hence inelastic. The predictions in the $\{m_{\rm DM}, \delta\}$-plane are given in figure~\ref{fig1} (see caption for the labeling), following the Bayesian inference procedure explained in~\cite{Arina:2011cu}. Scalar DM in the mass range above 2 TeV is excluded as explanation of DAMA annual modulation and CRESST-II excess by XENON100 and KIMS. We therefore do not consider any further the scalar case for asymmetric DM. Contrary fermionic DM is compatible at 99\% C.L. with all exclusion bounds in the mass range $50  \to  350$ GeV. This candidate is compatible as well with the CRESST commissioning run on W~\cite{Angloher:2008jj,Arina:2012fb}. Notice that at those masses the regions that explain DAMA and CRESST-II overlap.  For the fermionic candidate to have a mass splitting $\delta \simeq 150$ keV we require $f_\psi \geq 10^{-4}$. This hierarchy between $f_L$ and $f_\psi$, and consequently in the Majorana masses will fix the phenomenology of leptogenesis and DM asymmetry generation, as it will be discussed in~\ref{sec:asym}.

\section{Inflation from triplet scalar and Higgs boson}\label{sec:infl}

By adding non-minimal couplings to gravity for the Higgs and triplet, the action in the Jordan frame is given by:
\begin{equation}
S_{J}  =  \int {\rm d}^4 x \,\sqrt{-g}\,  \left[ \frac{R}{2}
 +   \left(\xi_H H^{\dagger} H + \xi_{\Delta} \Delta^{\dagger}\Delta + c.c.\right) \ R 
 -  |\mathcal{D}_{\mu} H|^2 - |\mathcal{D}_{\mu} \Delta|^2 - V_J(H,\Delta)\right] \,,
\label{eq:Sj}
\end{equation}
with the scalar content of the model described by the potential in equation~\ref{eq:ScalarPotential}. In order to have the usual interaction for gravity, it is usual to perform a conformal transformation from the Jordan frame to the Einstein frame, where the matter content ends up with non standard kinetic terms. The Higgs and triplet fields are defined in the unitarity gauge, for a total of three degree of freedom $h, \delta$ and $\theta$ that can be redefined in the large field limit as $r=\delta/h, \varphi = \sqrt{3/2}\log \left(1+\xi_h h^2 + \xi_\Delta \delta^2\right)$ and $\theta$. During inflation the mass value for the field $r$ is always large compared to the Hubble parameter, hence this heavy field sets quickly in a minimum of the potential: we are left with an effective theory for the light inflatons $\varphi$ and $\theta$. Note that with respect to the quadratic term in the scalar potential the mass terms are negligible as far as $M_\Delta \leq 10^{12}$ GeV, while the $\mu_H$ term becomes negligible for $\mu_H< 10^{10}$ GeV. Hence only the quartic term contributes to inflation and is $\theta$ independent, namely this model is equivalent to Higgs inflation:
\begin{equation}\label{eq:HiggsPot}
V(\varphi) = V_0\,  \left(1-e^{-2\varphi/\sqrt{6}}\right)^2 \,, 
\end{equation}
with
\begin{equation}
V_{\varphi \rm{-indep}} =
\left\{
\begin{array}{l}
V_0^{\rm (mixed)} =\frac{\lambda_\Delta \lambda_H - \lambda_{H\Delta}^2}{8\,  (\lambda_\Delta\, \xi^2_H + \lambda_H\, \xi^2_\Delta -2\lambda_{H\Delta} \xi_\Delta \xi_H)}\,, \\
V_0^{(H)} = \frac{\lambda_H}{8\xi_H^2} \,,\\
 V_0^{(\Delta)} = \frac{\lambda_\Delta}{8\xi_\Delta^2} \, .
\end{array}
\right.
\end{equation}
The three different $V_0$ arise from different minimum in which $r$ sits: depending on its value, the inflaton is an admixture of Higgs and triplet or a pure Higgs or a pure triplet. In figure~\ref{fig2} we show as example of the model parameter space for a mixed inflaton, constrained by matching the amplitude of the power spectrum with the WMAP data. The parameters are fixed at EW and run up to unitarity scale using the appropriate renormalization group equations, with the inclusion of the triplet and DM contributions. Since the triplet is at a scale of $10^8$ GeV, the inflationary constraints loosely bound the EW parameters, while $\lambda_\Delta$ or $\lambda_{H\Delta}$ are unaccessible at the low energy of the Large Hadron Collider (LHC).

\section{Generation of asymmetries in the visible and dark sectors}\label{sec:asym}

\begin{figure}[t]
\centering
\includegraphics[width=.8\columnwidth]{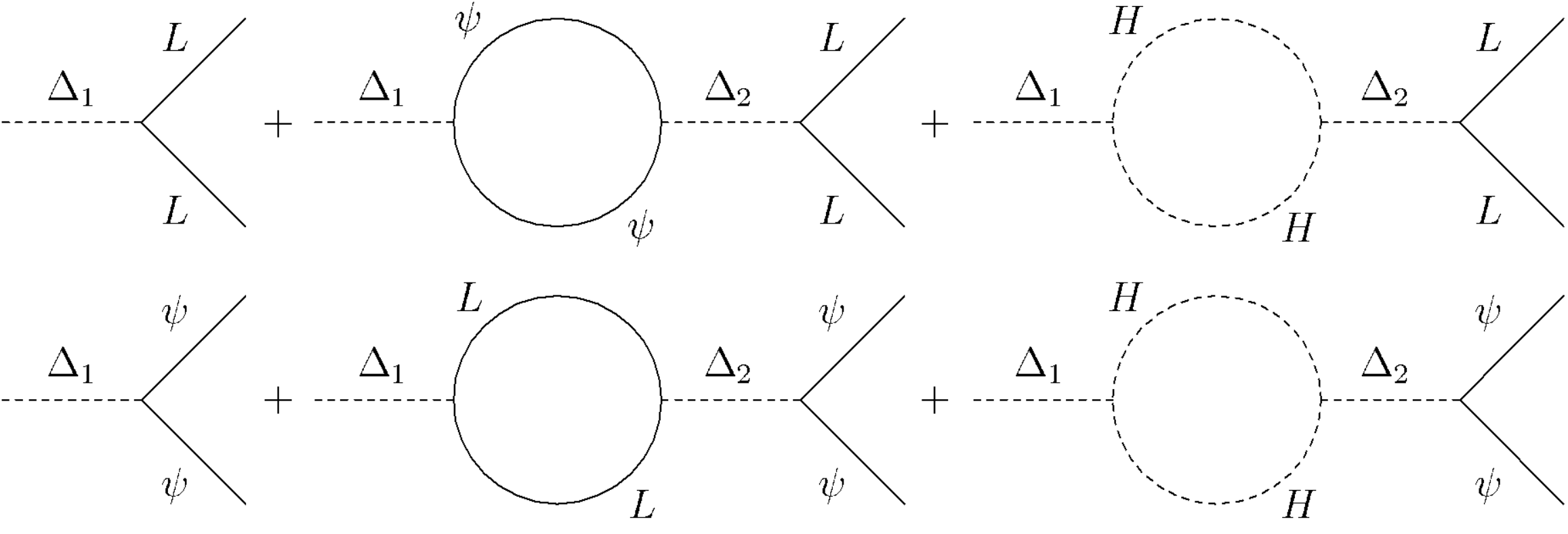}
\caption{The CP asymmetry in the leptonic and dark sectors, first and second raw respectively, generated by the interference of tree and one-loop self energy correction diagrams of the scalar triplet $\Delta_1$ and $\Delta_2$.}
\label{fig3}
\end{figure} 

From equations~\ref{eq:Lag-CP} and~\ref{eq:Lag-DM}, note that there are three channels for triplet decay: $\Delta \to HH$, $\Delta \to LL$ and $\Delta \to \psi\psi$. Since the couplings are complex, the quasi-equilibrium decay of $\Delta$ via these channels produce asymmetries in lepton and DM sectors, as described in figure~\ref{fig3}, with the requirement of having at least two scalar triplet $\Delta_1$ and $\Delta_2$. When the total decay rate of the lightest of the triplet mass eigenstate ($\zeta_1$) fails to compete with the Hubble expansion rate of the universe, then $\zeta_1$ decays and produces the asymmetries. These are proportional to the CP asymmetries $\epsilon_i$ (with $i=DM,H,L$), to the triplet yield $X_{\zeta_1}=n_{\zeta_1}/s$ ($s$ is the entropy density) and to the efficiency factor $\eta_i$, that account for the washout effect on the asymmetries due to the number violating processes involving $\psi$, $L$ and $H$. At a temperature above EW phase transition a part of the lepton asymmetry gets converted to the baryon asymmetry via the $SU(2)_L$ sphaleron processes. We solve the relevant Boltzmann equations for the quasi-equilibrium decays of $\zeta_1$ and require that the relic abundance of the visible and dark matter are satisfied:
\begin{equation}\label{eq:IMP}
\frac{\Omega_{\rm DM}}{\Omega_B} = \frac{1}{0.55 }\frac{m_{\rm DM}}{m_p} \frac{\epsilon_{\rm DM}}{\epsilon_L}
\frac{\eta_{\rm DM}}{\eta_L}\,,
\end{equation}
with $m_p$ being the proton mass, as well as the asymmetry in the baryon sector matches the measured ones:
\begin{equation}\label{eq:basym}
\frac{n_B}{n_\gamma} \sim 6.15 \times 10^{-10} = 7.02 \times 0.55 \times Y_L\,,
\end{equation}
where $Y_L$ is the yield for the leptons sector, defined as $Y_L = \eta_L \epsilon_L X_\zeta$. The sampling over the parameter space of the model, given by the branching ratios $B_L,B_{\rm DM}$, the CP asymmetries $\epsilon_L$ and $\epsilon_{\rm DM}$ and the dark matter mass $m_{\rm DM}$ is performed via MCMC methods. 
\begin{figure}[t]
\begin{minipage}[t]{0.5\textwidth}
\centering
\includegraphics[width=1.\columnwidth,trim=15mm 25mm 10mm 15mm, clip]{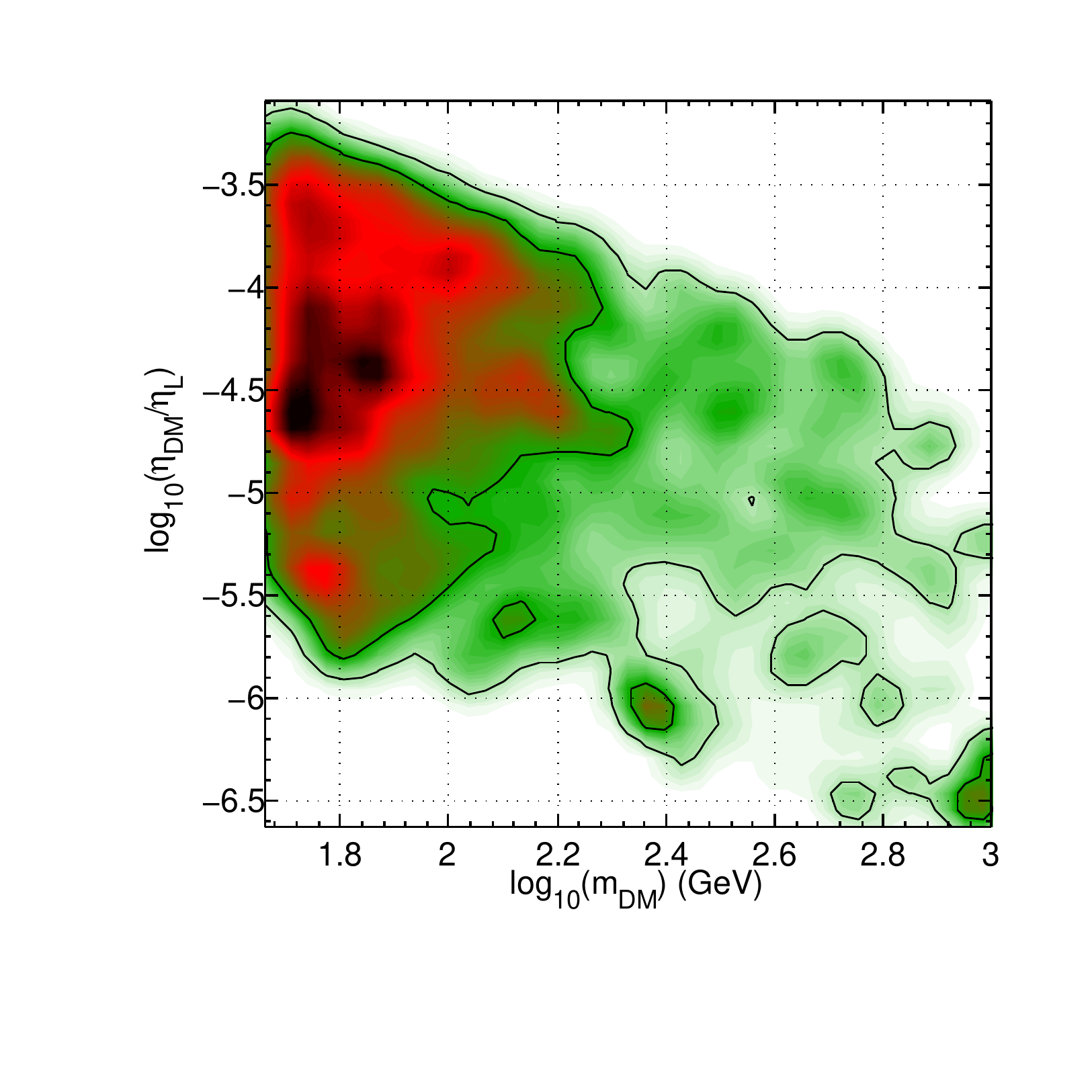}
\caption{2D credible regions at 68\% and 95\% C.L. in the \{$m_{\rm DM},\eta_{\rm DM}/\eta_{L}$\}-plane. All of other parameters have been marginalized over.\label{fig4}}
\end{minipage}
\hspace*{0.2cm}
\begin{minipage}[t]{0.5\textwidth}
\centering
\includegraphics[width=1.\columnwidth,trim=15mm 25mm 10mm 15mm, clip]{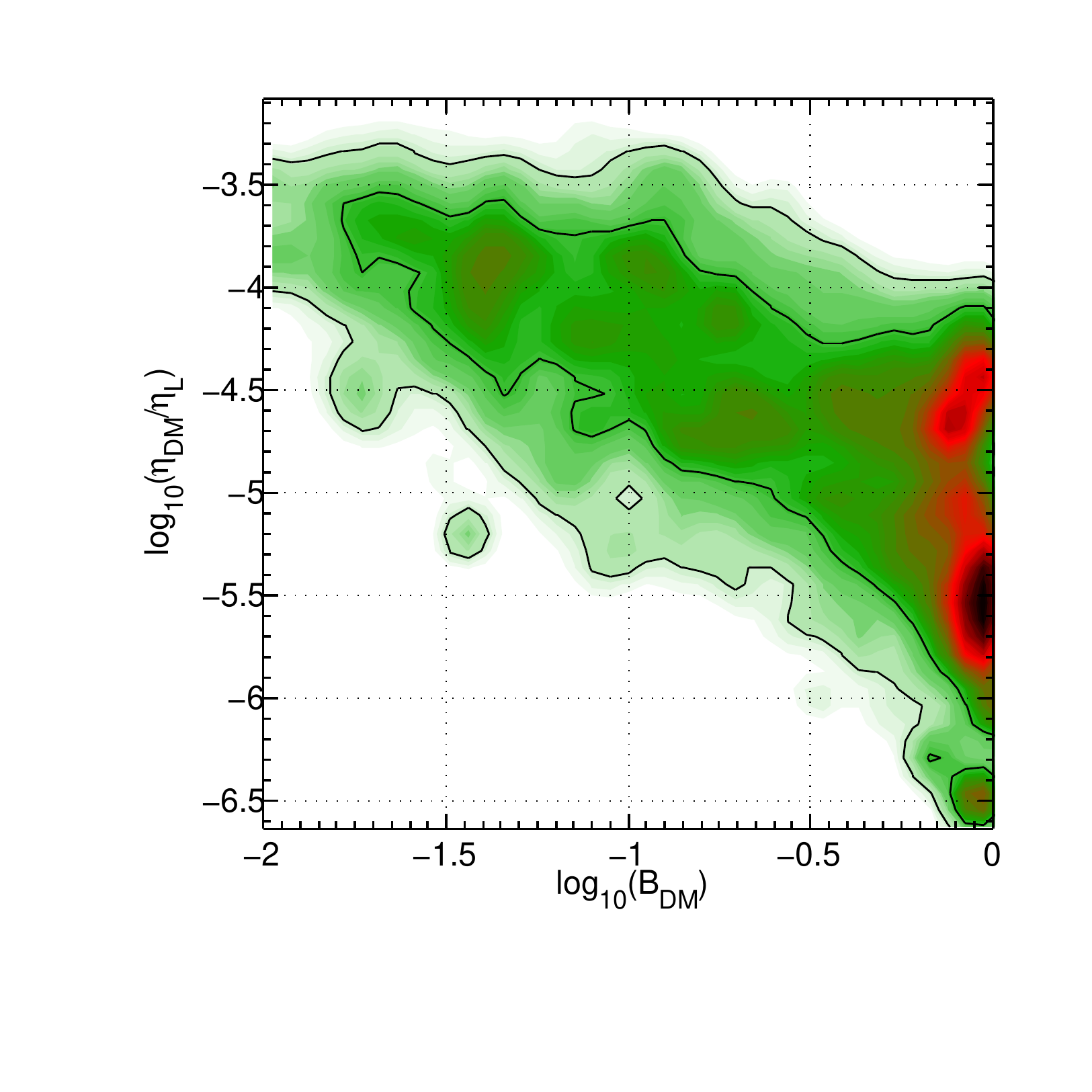}
\caption{Same as left in the \{$B_{\rm DM},\eta_{\rm DM}/\eta_{L}$\}-plane. For details about the MCMC sampling we refer to~\cite{Arina:2012fb}.\label{fig5}}
\end{minipage}
\end{figure}

An overview of the results is shown in figure~\ref{fig4} and figure~\ref{fig5}, where the 2D posterior pdfs are plotted for different parameters of the model. From figure~\ref{fig4} we see that all the mass range from 45 GeV up to 1 TeV can lead to successful leptogenesis and  to an asymmetric dark matter candidate, namely satisfies equations~\ref{eq:basym} and~\ref{eq:IMP}. Note from the posterior pdf that the most favored region is for low mass candidates, even though there are candidates viable up to 1 TeV with smaller statistical significance. For a DM mass up to around 150 GeV, the preferred values of the ratio $\eta_{\rm DM}/\eta_L$ are of  $\mathcal{O}(10^{-4}-10^{-5})$, which compensate the large CP asymmetry ratio $\epsilon_{\rm DM}/\epsilon_L$. For DM masses around ${\mathcal O}$(TeV), $\epsilon_{\rm DM}/\epsilon_L$ is even larger for the preferred values of $\eta_{\rm DM}/\eta_L$, which decreases down to $10^{-6}$. For a triplet mass of $10^8$ GeV the important quantities which drive Boltzmann equations are the branching ratios. In figure~\ref{fig5} we show the correlation of $\eta_{\rm DM}/\eta_L$ versus $B_{\rm DM}$ with the 68\% and 95 \% credible regions. We see that the largest efficiency ratio $\eta_{\rm DM}/\eta_L$ is preferred when $B_{\rm DM} \to 1$, which implies a small $B_{\rm L} \to 10^{-3} - 10^{-4}$. This is because of the required hierarchy between the sub-eV neutrino mass and the Majorana mass splitting between the DM mass eigenstates, mentioned in section~\ref{sec:model}. In all cases the slow channel that builds and conserves the asymmetry is the leptonic one. The fast channel is either the Higgs, to compensate the neutrino mass in the total rate (see~\cite{Hambye:2005tk,Chun:2006sp} for accurate discussion) or the DM one. Since the DM channel is not related to the neutrino mass via the total rate, its branching ratio can assume different values all along the DM mass range. The small values for the efficiency ratio are compensated by the large CP asymmetry ratio, as confirmed in figure~\ref{fig5}. We note that without this constraint on the hierarchy between Majorana masses the behavior for the CP asymmetry ratio and for the efficiency factor ratio  would be the opposite ones, as shown in~\cite{Arina:2011cu}.

\section{Conclusions}

In these Proceedings we have discussed an extension of the SM with two heavy triplet scalars whose partial decay to SM leptons and inert doublet scalars ($\chi$) or vector like fermions ($\psi$), could explain a common origin of asymmetric dark matter and visible matter through the leptogenesis mechanism. The induced vev of the triplets give rise to neutrino masses, as required by the oscillation experiments, via the type-II seesaw mechanism. 

Regarding the DM phenomenology, odd under a $Z_2$ symmetry and thus stable, for the scalar candidate fast oscillations between $\chi_0$ and $\overline{\chi_0}$ strongly deplete the asymmetry below EW symmetry breaking scale. Therefore, the survival of the asymmetry leads to a lower bound on the DM mass of $2$ TeV so that the DM freezes out before it begins to oscillate. Both scalar and fermionic candidate scatters inelastically off nuclei in underground detectors. By performing a Bayesian analysis we found that
an asymmetric scalars of mass larger than 2 TeV is excluded at 90\% C.L. by the XENON100 data
while an asymmetric fermionic DM of mass ${\mathcal O}(100)$ GeV is suitable to explain DAMA annual modulation
signal within the 99\% C.L. given by the XENON100 experiment. For the first time we propose it as explanation of the CRESST-II excess as well.

The indication at the LHC of a SM like Higgs boson with mass around 125-126 GeV suggests that the SM vacuum might be metastable. The scalar triplet extension of the SM evades as well the possibility of having a vacuum instability at least up to the unitarity scale. We introduced non-minimal couplings to gravity for both scalar triplet and the SM Higgs. In presence of these couplings the scalar triplet, mixed with the SM Higgs, drives inflation in the early Universe. We showed that the extended scalar potential gives rise to slow-roll single field inflation, once the heavy field is stabilized at a minimum of the potential. In general the inflaton is an admixture of triplet scalar and the SM Higgs or given by a pure Higgs or triplet field. Unfortunately it is not possible to measure the quartic couplings of the triplet at the LHC, because of its large mass. Hence it is not possible to distinguish between the three type of inflationary pictures and standard Higgs inflation. We leave for future work numerical treatment of Higgs inflation and its variant as well as future investigation on the triplet model related to Higgs properties, as soon as more data at LHC will be collected.

\ack
We thank Jinn-Ouk Gong for useful discussion. This work acknowledges use of the COSMO computing resource at CP3 of Louvain University.

\section*{References}
\bibliographystyle{iopart-num.bst}
\bibliography{biblio}

\end{document}